\documentclass[%
reprint,
showpacs,preprintnumbers,
amsmath,amssymb,
longbibliography
]{revtex4-1}

\usepackage{graphicx}
\usepackage{dcolumn}
\usepackage{bm}
\usepackage{colortbl}
\usepackage[unicode=true,colorlinks=true,citecolor=blue,urlcolor=blue]{hyperref}
\usepackage{braket}

\newcommand{\e}{\mathrm{e}}
\renewcommand{\i}{{\rm i}}
\renewcommand{\d}{{\rm d}}
\newcommand{\Tr}{\mathop{\rm Tr}}
\newcommand{\arctg}{\mathop{\rm arctg}}
\newcommand{\Si}{\mathop{\rm Si}}
\newcommand{\Ci}{\mathop{\rm Ci}}
\renewcommand{\ni}{{\bar\imath}}

\begin{document}


\title{Universal power law decay of spin polarization in double quantum dot}

\author{V.\,N.\,Mantsevich}\email{vmantsev@gmail.com}
\affiliation{Chair of semiconductors and cryoelectronics and Quantum technology center, Faculty of
Physics, Lomonosov Moscow State University, 119991 Moscow, Russia}

\author{D.\,S.\,Smirnov}\email{smirnov@mail.ioffe.ru}
\affiliation{Ioffe Institute, 194021, St. Petersburg, Russia}

\begin{abstract}
We study the spin dynamics and spin noise in a double quantum dot taking into account the interplay between hopping, exchange interaction and the hyperfine interaction. At short time scales the spin relaxation is governed by the spin dephasing in the random nuclear fields. At long time scales the spin polarization obeys universal power law $1/t$ independent of the relation between all the parameters of the system. This effect is related to the competition between the spin blockade effect and the hyperfine interaction. The spin noise spectrum of the system universally diverges as $\ln(1/\omega)$ at low frequencies. 
\end{abstract}

 \keywords{} \maketitle

\section{Introduction}

The most fascinating discoveries in the solid state physics in the XXI$^{\mbox{st}}$ century are related to the spin degree of freedom of electrons. Intense studies of the spin-related phenomena led to the formation of a new branch in the solid state physics -- spintronics~\cite{dyakonov_book}. The spin related phenomena are most pronounced in the low dimensional structures due to the enhanced role of the spin-orbit and hyperfine interactions~\cite{manchon2015new,ganichev2012spin,Hopping_spin}. From practical point of view, the most promising for quantum information processing are zero-dimensional nanosystems, such as shallow impurities, color centers and quantum dots (QDs).

There are two complementary approaches to study spin-related phenomena in the QDs. The first one is based on the optical spin orientation, manipulation and detection, and is usually applied to the self-organized quantum dots~\cite{press08,berezovsky08,A.Greilich07212006}. The second one is based on the electrical spin injection and detection in the gate-defined quantum dots~\cite{hanson07}, which makes use of the external magnetic field. An interesting and promising system for the latter approach is a double quantum dot~\cite{petta05,Bluhm2011}, which demonstrates the Pauli or spin blockade effect~\cite{Ono1313,PhysRevB.72.165308}. This effect was studied in detail theoretically~\cite{Coish2005,Fransson2006,PhysRevLett.96.176804,Taylor2007,Mantsevich_ssc_2017,Mantsevich_prb_2018}, but the spin dynamics was investigated mainly in the presence of electric current and external magnetic field.

In this work we study manifestations of the spin blockade effect in the spin dynamics of double quantum dot isolated from the environment in the absence of external magnetic field. Our theory can be also applied for an isolated pair of donors, which are close to each other, but far enough from the other donors.


The spin dynamics in quantum dots in zero magnetic field is largely driven by the hyperfine interaction with the host lattice nuclear spins~\cite{book_Glazov}. In a double quantum dot the exchange interaction~\cite{KKavokin-review,noise-exchange-eng} and electron hopping~\cite{PhysRevB.90.201203,Glazov_hopping} are also important and affect the spin dynamics. We stress, that we consider only hopping between the QDs, but not to the contacts or substrate~\cite{maslova2019probing,Mantsevich201433}. The interplay between exchange interaction, hopping and hyperfine interaction can be hardly investigated for the large spin ensembles. But the double quantum dot system considered here allows for the exact solution and gives some hints about spin dynamics in larger spin systems.

In our study we focus on two effects: the spin relaxation and spin noise. The first one assumes the spin orientation and measurement of the spin polarization decay. The second one is based on the continuous measurement of the dynamics of spin fluctuations in the thermal equilibrium~\cite{Zapasskii:13,Oestreich-review}. We demonstrate, that in both cases the spin dynamics essentially consists of the spin precession in a random nuclear field and slow power law relaxation. The latter effect is a consequence of the interplay between the spin blockade and the hyperfine interaction.

The paper is organized as follows. In the next section we present the model of the system under study. In Sec.~\ref{sec:general} we present our approach to calculate the spin dynamics and spin noise spectrum. We show numerical results for the arbitrary relation between the system parameters and stress the universality of the power law $\propto1/t$ spin relaxation. Then in Sec.~\ref{sec:limits} we derive the analytical results in the limiting cases, which explain the numerical results. Further, in Sec.~\ref{sec:discussion} we discuss the limits of applicability of our model, and finally summarize our findings in Sec.~\ref{sec:conclusion}.

\newpage
\section{Model}
\label{sec:model}

\begin{figure}
  \centering\includegraphics[width=0.8\linewidth]{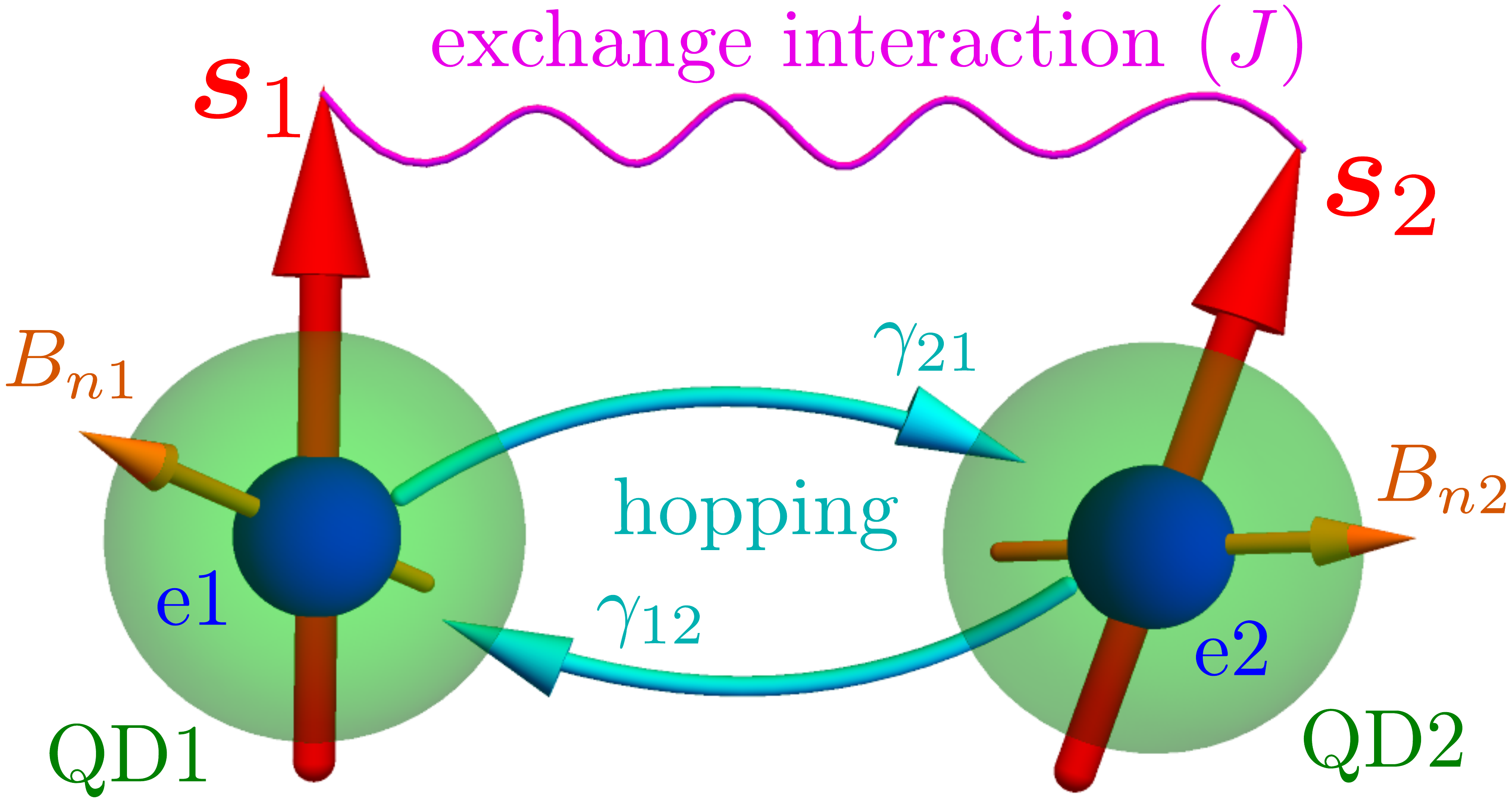}
\caption{Sketch of the double QD system. The two electrons are represented by the blue balls, and their spins --- by the red arrows. Orange arrows show the Overhauser field acting in each dot, magenta spring denotes the exchange interaction. The navy arrows show the possible hops of electrons between the QDs (green transparent balls).}
\label{fig:sketch}
\end{figure}

We consider a double QD with two electrons, as shown in Fig.~\ref{fig:sketch}. We assume that the two electrons can be localized either in different or in the same QD, and can hop between the QDs. We take into account the exchange interaction between electrons and their hyperfine interaction with the host lattice nuclear spins. The Hilbert space of the system under study consists of six states: the two singlet states, when the two electrons are localized in the same QD, plus another singlet state and three triplet states, when the two electrons are localized in different QDs.

The Hamiltonian of the system has the form:
\begin{equation}
  \mathcal H=\sum_{i,\sigma}E_in_i^\sigma+\sum_{i}U_in_{i}^{+}n_{i}^{-}+J\textbf{s}_1\textbf{s}_2+\hbar\sum_{i}\bm \Omega_{i}\textbf{s}_{i}.
\label{eq:H_e}
\end{equation}
Here in the first term $E_i$ are the localization energies of electrons in the $i$th QD ($i=1,2$) and $n_i^\sigma$ ($\sigma=\pm$) are the occupancies of the states, characterized by the spin index $\sigma$. The corresponding operators can be written using the Fermi creation (annihilation) operators $c_{i,\sigma}^{\dag}$ ($c_i^\sigma$) as $n_i^\sigma=c_{i,\sigma}^\dag c_{i,\sigma}$. The second term in Eq.~\eqref{eq:H_e} describes the on-site electron repulsion with the Hubbard energy $U_i$. The third term is the exchange interaction, characterized by the constant $J$. The spin operators can be expressed as
\begin{equation}
  \bm s_i=\frac{1}{2}\bm\sigma_{\sigma\sigma'}c_{i\sigma}^\dag c_{i\sigma'},
\end{equation}
where $\bm \sigma=(\sigma_x,\sigma_y,\sigma_z)$ is the vector composed of the Pauli matrices. Finally, the last term in Eq.~\eqref{eq:H_e} is the hyperfine interaction, where $\bm\Omega_i$ is the spin precession frequency in the fluctuation of host lattice nuclear spin polarization. In this study we assume the number of host lattice nuclear spins to be large, so that the Overhauser field can be considered as static (``frozen'')~\cite{merkulov02}.

The electron hopping being inelastic process, can not be described be electron Hamiltonian solely. Therefore one has to consider the total Hamiltonian
\begin{equation}
  \mathcal H_{tot} = \mathcal H + \mathcal H_{ph}+\mathcal V,
\end{equation}
which takes into account a phonon Hamiltonian $\mathcal H_{ph}$ and an electron phonon interaction $\mathcal V$. The phonon energy is given by
\begin{equation}
  \mathcal H_{ph}=\hbar\sum_{\bm q}\Omega_{q}b_{\bm q}^\dag b_{\bm q},
\end{equation}
where  $\Omega_{\bm q}$ is the phonon frequency, corresponding to the wavevector $\bm q$, and $b_{\bm q}^\dag$ ($b_{\bm q}$) is the phonons creation (annihilation) operator. We assume the phonon polarization index to be included in $\bm q$. The electron-phonon interaction after the canonical (polaron) transformation~\cite{Firsov-book,BryksinReview} reads
\begin{multline}
  \mathcal V=t\sum_\sigma
\exp\left\{-\sum_{\bm q}\gamma_q\left[\left(\e^{\i\bm{qR}_1}-\e^{\i\bm{qR}_2}\right)b_{\bm q}-\rm{h.c}\right]\right\}
\\\times
c_{1\sigma}^\dag c_{2\sigma}
+\rm{h.c.}\:,
\end{multline}
where $t$ is the hopping constant, $\gamma_q=v_q/(\hbar\Omega_q)$ with $v_q$ being the electron-phonon interaction constant, and $\bm R_{1,2}$ being the coordinates of the QDs.

The spin dynamics in the system can be described using the density matrix formalism. In the description of electron hopping we assume that the first two terms in the Hamiltonian~\eqref{eq:H_e} and the temperature exceed by far the two latter terms, so the states, where two electrons are in the different QDs, have nearly the same energy. In this case the off diagonal matrix elements between the states with the essentially different energy can be neglected. As a result the system is described by the $4\times4$ density matrix $\rho$ in the basis of the four states of two electrons in different QDs, and the two probabilities $P_{i}$ to find the two electrons in the QD $i$.

In the two lowest orders in the hopping amplitude ($t$) the total density matrix of the electron system $\rho_{tot}$ satisfies the equation
\begin{multline}
  \label{eq:new}
  \dot{\rho}_{tot}=-\frac{\i}{\hbar}[\mathcal H,\rho_{tot}]\\+\frac{\pi}{\hbar}\left\langle\left[2{\mathcal V}\rho_{tot}{\mathcal V}
    -\rho_{tot}{\mathcal V}^2-{\mathcal V}^2\rho_{tot}\right]\delta(E_i-E_f)\right\rangle_{ph},
\end{multline}
where the first line describes the coherent spin dynamics and the second one --- the electron hopping. The angular brackets denote averaging of the phonon creation and annihilation operators over the phonon states. The energies $E_{i}$ and $E_f$ are the total energies of the system before and after the hop, respectively, including the phonon energy. Note that we do not include $\mathcal V$ in $\mathcal H$, assuming it to be negligible in comparison with the other terms.

From Eq.~\eqref{eq:new} we find, that the electron density matrix $\rho$ obeys the master equation
\begin{multline}
  \label{eq:rho}
  \dot{\rho}=-\frac{\i}{\hbar}\left[\mathcal H,\rho\right]+\frac{1}{2}\sum_{i}\sum_{\sigma,\sigma'}\left[2\Gamma_{i\ni}c_{i\sigma}^\dag c_{\ni\sigma}P_{\ni}c_{\ni\sigma'}^\dag c_{i\sigma'}
  \right.\\
  \left.-\gamma_{i\ni}\left(\rho c_{\ni\sigma'}^\dag c_{i\sigma'}c_{i\sigma}^\dag c_{\ni\sigma}+c_{\ni\sigma'}^\dag c_{i\sigma'}c_{i\sigma}^\dag c_{\ni\sigma}\rho\right)\right],
\end{multline}
where the symbol $\ni$ denotes the other quantum dot than $i$ ($\ni=2$ if $i=1$ and $\ni=1$ if $i=2$) and we introduce the rates $\gamma_{i\ni}$ and $\Gamma_{i\ni}$ describing the hopping from QD $\ni$ to $i$ when the QD $i$ is occupied or empty, respectively. In the second order in the electron-phonon interaction ($\gamma_q$) the hopping rate with the change of energy by $\Delta E$ is~\cite{Efros89_eng,Hopping_spin}
\begin{equation}
  \gamma(\Delta E)=\frac{2\pi}{\hbar}t^22\gamma_{q_{\Delta E}}^2D(\Delta E)\left[N_{\Delta E}+\theta(-\Delta E)\right],
\end{equation}
where $q_{\Delta E}$ is the phonon wave vector corresponding to a phonon
with the energy $\Delta E$, $D(\Delta E)$ stands for the density of phonon states, $N_{\Delta E}=1/\left[\exp(\Delta E/k_B T)-1\right]$
is the occupancy of the corresponding states with $T$ being the temperature, and $\theta(x)$ is the Heaviside step function. In the higher orders in the electron-phonon interaction the expression for the hopping rate is different, but its explicit form is unimportant for our study. The specific hopping rates between the QDs are given by
\begin{subequations}
  \begin{equation}
    \gamma_{i\ni}=\gamma(E_i-E_\ni+U_i),
  \end{equation}
  \begin{equation}
    \Gamma_{i\ni}=\gamma(E_i-E_\ni-U_i).
  \end{equation}
\end{subequations}
One can see, that in the general case the relation $\Gamma_{i\ni}\ge\gamma_{i\ni}$ holds, because of the electron Coulomb repulsion in the same QD.

Similarly to Eq.~\eqref{eq:rho} the probabilities $P_{i}$ obey
\begin{multline}
  \label{eq:P}
  \dot{P}_{i}=\frac{1}{2}\sum_{\sigma,\sigma'}\left[2\gamma_{i\ni}c_{i\sigma}^\dag c_{\sigma}\rho c_{\ni\sigma'}^\dag c_{i\sigma'}
  \right.\\\left.
    -\Gamma_{\ni i}\left(P_{i}c_{i\sigma'}^\dag c_{\ni \sigma'}c_{\ni \sigma}^\dag c_{i\sigma}+c_{i\sigma'}^\dag c_{\ni\sigma'}c_{\ni\sigma}^\dag c_{i\sigma}P_{i}\right)\right].
\end{multline}
The electron conservation rule for this system can be written as
\begin{equation}
  \label{eq:1}
  P_{1}+P_{2}+P_{12}=1
\end{equation}
where $P_{12}=(n_1^++n_1^-)(n_2^++n_2^-)=\Tr\rho$ is the probability to find the two electrons in the different QDs.

We recall that we assume the nuclear fields $\bm\Omega_i$ to be frozen. They are created by the nuclear spin fluctuations and are described by the Gaussian distribution function
\begin{equation}
  \label{eq:F}
  \mathcal
  F(\bm\Omega_i)=\frac{1}{(\sqrt{\pi}\delta)^3}e^{-\Omega_i^2/\delta^2},
\end{equation}
with the parameter $\delta$ characterizing the dispersion. In order to obtain experimentally observable spin dynamics, the solution of the spin dynamics equations should be averaged over this distribution function. In the next section we demonstrate, that this procedure ultimately leads to the power low spin decay $\propto 1/t$ at long time scales.

\section{Spin relaxation and spin noise}
\label{sec:general}

The master Eq.~\eqref{eq:rho} can be rewritten in the form of equations for the spin operators $\bm s_{i}$ and their correlation functions $s_i^\alpha s_j^\beta$, where $\alpha$ and $\beta$ are the Cartesian indices. Their average values can be expressed through the density matrix as $\braket{\bm s_i}=\Tr(\bm s_i\rho)$ and $\braket{s_i^\alpha s_j^\beta}=\Tr(s_i^\alpha s_j^\beta\rho)$.

The electron spins obey
\begin{subequations}
  \label{system_spin}
  \begin{equation}
    \label{eq:spin_i}
    \frac{\d\bm s_{i}}{\d t}=\bm \Omega_{i}\times \bm s_{i}+\frac{J}{\hbar}\bm s_{\ni}\times \bm s_{i}-\frac{\gamma}{2}(\bm s_{i}-\bm s_{\ni}),
  \end{equation}
where $\gamma=\gamma_{12}+\gamma_{21}$ is the total hopping rate for the singlet state of the two electrons in different QDs. The first term on the right hand side in this equation describes the spin precession with the frequency $\bm\Omega_{i}$. Similarly the second term describes the electron spin precession in the effective exchange magnetic field of another electron. Finally, the last term describes the electron hopping. One can see, that this term vanishes in the case of $\bm s_1=\bm s_2$ due to the spin blockade. By contrast, the hopping rate equals to $\gamma$ when the two electrons are in the singlet state, i.e. $\bm s_1=-\bm s_2$. We recall, that the relation $k_BT,U_{1,2}\gg J,\hbar\delta$ is assumed, so there is no spin polarization in the thermal equilibrium.

We stress, that the term $\bm s_1\times\bm s_2$ in Eq.~\eqref{eq:spin_i} does not simply reduce to the product of the two average values, but should be treated as a vector composed of the spin correlators. The spin correlation functions in general obey the equations
  \begin{widetext}
  \begin{equation}
    \label{eq:ss}
    \frac{\d}{\d t}\left(s_1^\alpha s_2^\beta\right)=\varepsilon_{\alpha\gamma\delta}\Omega_1^\gamma s_1^\delta s_2^\beta+\varepsilon_{\beta\gamma\delta}\Omega_2^\gamma s_1^\alpha s_2^\delta+\frac{J}{4\hbar}\varepsilon_{\alpha\beta\gamma}\left(s_1^\gamma-s_2^\gamma\right)-\frac{\gamma}{2}\left(s_1^\alpha s_2^\beta-s_1^\beta s_2^\alpha\right)+\frac{\delta_{\alpha\beta}}{2}\left(\gamma P_s-\Gamma_{21}P_{1}-\Gamma_{12}P_{2}\right),
  \end{equation}
  \end{widetext}
\end{subequations}
where $P_s=P_{12}/4-\bm s_1\bm s_2$ is the occupancy of the singlet state in the two different QDs. Explicit form of these equations is given in Appendix~\ref{app:eqs}. The first two terms in the right hand side of this expression describe the spin precession in the nuclear field. The third term is related to the exchange interaction and reduces to the first power of spin operators for the spin one half particles. The rest of the terms describe the hopping of electrons and deserve a longer discussion.

The hopping of two electrons to the same QD brings the system to the singlet state with the zero total angular momentum. In the same time, the hopping does not change the total angular momentum, so it is allowed only for the singlet spin state in agreement with the Pauli exclusion principle. The correlators $s_1^\alpha s_2^\beta$ can be combined in the groups, which transform according to the representations $D_2$, $D_1$ and $D_0$ of SU(3) group. The five correlators $s_1^\alpha s_2^\beta+s_1^\beta s_2^\alpha$ with $\alpha\neq\beta$ belong to $D_2$ representation and do not decay, because they require the two electron spins to be parallel. The three combinations $s_1^\alpha s_2^\beta-s_1^\beta s_2^\alpha$ belong to $D_1$ representation and decay with the rate $\gamma$, which is described by the penultimate term in Eq.~\eqref{eq:ss}. Finally, the correlator $\bm s_1\bm s_2$ belongs to $D_0$ representation, and it couples to the scalar occupancies $P_{1}$, $P_{2}$ and $P_{12}$, which is described by the last term with the square brackets. This term consists of two contributions: hopping to the states, where the two electrons are localized in the same QD with the rate $\gamma$, and hopping from these states with the rates $\Gamma_{12}$ and $\Gamma_{21}$.

The above equations describe the spin dynamics and should be accompanied by kinetic equations for the occupancies of the states. Taking into account Eq.~\eqref{eq:1} it is enough to write the two equations
  \begin{equation}
    \label{eq:P2}
    \dot{P}_{i}=-2\Gamma_{\ni i}P_{i}+2\gamma_{i\ni}P_s,
  \end{equation}
The set of 18 Eqs.~\eqref{eq:1}, \eqref{system_spin}, and~\eqref{eq:P2} is equivalent to Eqs.~\eqref{eq:rho} and~\eqref{eq:P} and completely describes the spin and charge dynamics.

To simplify the following analysis we set all the hopping rates, $\Gamma_{12}$, $\Gamma_{21}$, $\gamma_{12}$ and $\gamma_{21}$, equal to $\gamma/2$. Physically this corresponds to the high temperature limit (the thermal energy much larger, than the Hubbard energies $U_i$). Then it is convenient to introduce the parameter
\begin{equation}
  X=P_1+P_2-2P_s,
\end{equation}
which describes the deviation of the occupancies from their steady state values. This parameter simply obeys
\begin{equation}
  \label{system_corr_3}
  \frac{\d X}{\d t}=2(\bm\Omega_1-\bm\Omega_2)(\bm s_1\times\bm s_2)-4\gamma X.
\end{equation}
Note that the same parameter describes also the dynamics of the spin correlators in Eq.~\eqref{eq:ss}. Thus in this case one can consider only 16 equations: Eqs.~\eqref{system_spin} and~\eqref{system_corr_3}. The spin dynamics can be calculated for the given initial conditions, and the double QD system is characterized in total by the three parameters: $J$, $\delta$ and $\gamma$.

To describe the spin relaxation we consider the initial conditions $\bm s_1(0)=\bm s_2(0)=\bm e_z/2$, where $\bm e_z$ is a unit vector along some $z$ axis. These initial conditions correspond to the optical spin orientation, and they are opposite to what is realized in electrically controlled double QD system~\cite{Taylor2007}. Fig.~\ref{figure2}(a) shows the evolution of the $z$ component of the total spin $\bm S=\bm s_1+\bm s_2$ in the most complicated case, when all the parameters are of the same order $J/\hbar\sim\delta\sim\gamma$ (their values are given in the figure caption). The spin dynamics can be separated into two contributions, below we describe them separately:

\begin{figure}
\includegraphics[width=\linewidth]{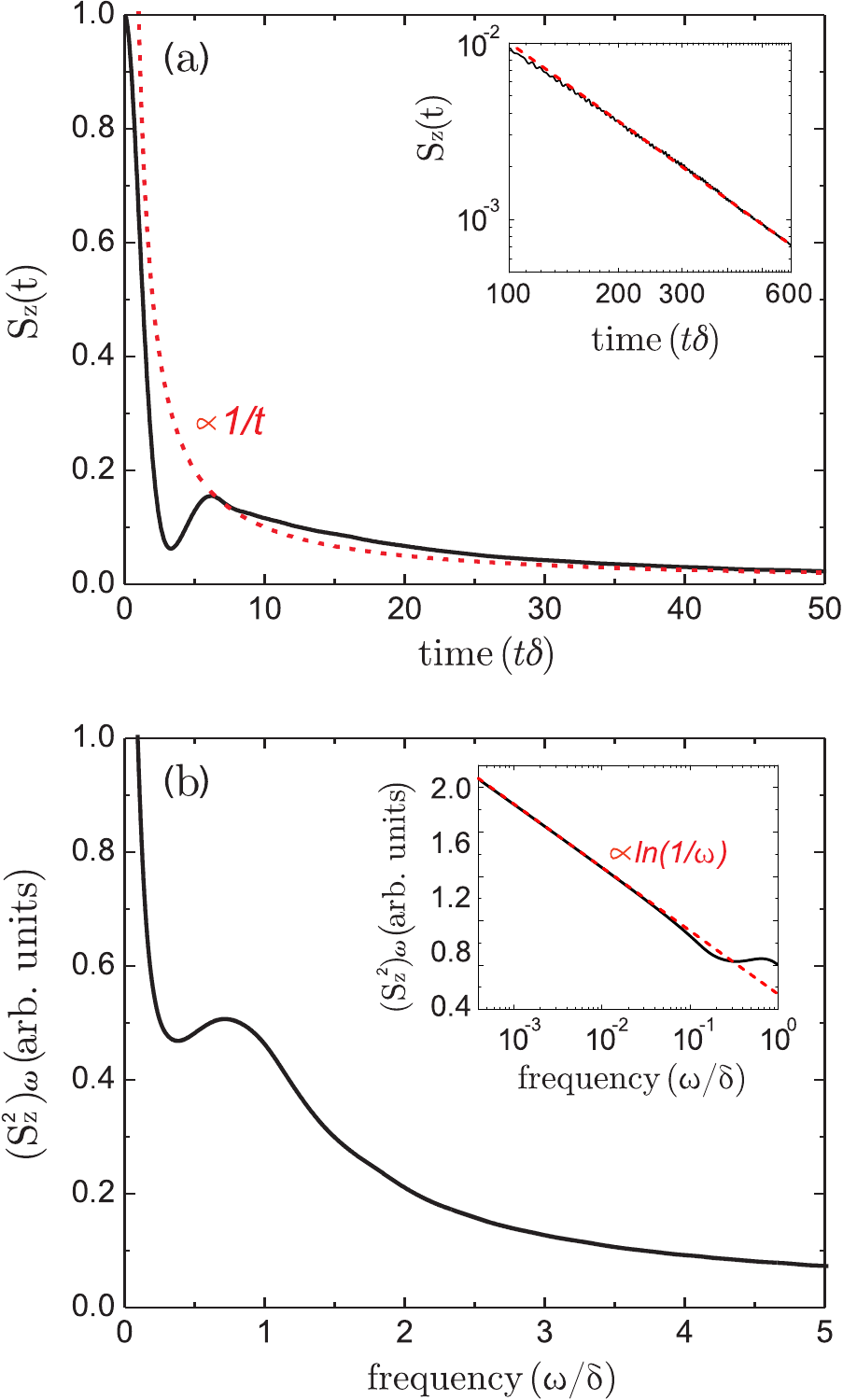}%
\caption{(a) The spin relaxation for the initial conditions $\bm s_i(0)=\bm e_z/2$. The red dashed lines show the asymptote $\propto 1/t$.  (b) The spin noise spectrum. The red dashed curve in the inset shows the asymptote $\propto\ln(1/\omega)$. The parameters of the calculation are $J=0.9\hbar\delta$ and $\gamma=1.1\delta$. 
} \label{figure2}
\end{figure}

(i) The total spin quickly decays from $1$ to less than $0.1$ and then increases again at $t\delta\sim 5$. This time dependence is typical for the spin dephasing in random Overhauser field~\cite{schulten,merkulov02,PhysRevLett.88.186802}. Notably, the spin polarization does not decay to zero due to the conservation of the spin component parallel to the Overhauser field in each QD. The exchange interaction ``exchanges'' the electrons in the two QDs, so the direction of precession of the given electron spin changes. This, however, also does not lead to the complete spin relaxation~\cite{noise-exchange-eng}. Indeed, in the limit of very strong exchange interaction the hyperfine field does not mix the singlet and triplet states, so the component of the total spin $\bm S$ along the average Overhauser field $\bm\Omega=(\bm\Omega_1+\bm\Omega_2)/2$ is conserved.

(ii) At the long timescales the total spin slowly decays to zero due to the \textit{hopping} of electrons between the QDs. In fact, this is a power law decay $S_z(t)\propto 1/t$, as shown by the red dashed line in Fig.~\ref{figure2}(a). This asymptotic is more clearly shown in the inset, where the longer time scales are shown in the bilogarithmic scale. We checked numerically, that this law of spin relaxation is valid for arbitrary relations between the parameters $J$, $\gamma$ and $\delta$. Moreover this law will be derived analytically in a number of limiting cases in the next section. To understand the effect qualitatively we note, that in the exceptional case $\bm\Omega_1\parallel\bm e_z$ and $\bm\Omega_2\parallel\bm e_z$ the total spin does not change, and the hopping is also forbidden for the initial condition $\bm s_1=\bm s_2$ [see Eqs.~\eqref{system_spin}]. So in this case the spin polarization (in our model) does not decay at all. In the more probable situation, when $\bm\Omega_1\parallel\bm\Omega_2$ the component of the total spin along this direction does not decay either because of the spin blockade. Finally, in the general case of arbitrary angle between $\bm\Omega_1$ and $\bm\Omega_2$ the spin polarization decays the longer the smaller is the angle. Averaging over the Gaussian distribution of the Overhauser fields results in the power law decay $S_z(t)\propto 1/t$ at long time scales.

The slow spin decay can be conveniently revealed in the frequency domain. Experimentally the spin dynamics at low frequencies can be studied by means of the spin noise spectroscopy~\cite{Zapasskii:13}. This method is based on the measurement of the correlation functions of the spin fluctuations in the thermal equilibrium.  The spin noise spectrum $(\delta S_z^2)_\omega$ is defined as a Fourier transform of the autocorrelation function
\begin{equation}
  \label{eq:spectrum}
  (\delta S_z^2)_\omega=\int\limits_{-\infty}^\infty \braket{\delta S_z(t)\delta S_z(t+\tau)}\e^{\i\omega\tau}\d\tau,
\end{equation}
where the angular brackets denote averaging over $t$. In the equilibrium the spin polarization is absent, so $\braket{S_z(t)}=0$ and $\delta\bm S=\bm S$ in the system under study.

To calculate the correlation functions we note, that the correlators at $\tau=0$ can be simply found from the steady state solution of the equations of motion. One finds, that the correlation functions of $S_z$ with all the other operators in Eqs.~\eqref{system_spin} and~\eqref{system_corr_3} are zero except for
\begin{equation}
  \braket{S_z s_i^z}=\frac{\braket{P_{12}}}{4}.
\end{equation}
In the thermal equilibrium $\braket{\bm s_1\bm s_2}=0$, so from Eqs.~\eqref{eq:P2} and Eq.~\eqref{eq:1} we find that
\begin{equation}
  \braket{P_{12}}=\left(1+\frac{\gamma_{12}}{4\Gamma_{21}}+\frac{\gamma_{21}}{4\Gamma_{12}}\right)^{-1}.
\end{equation}
In the high temperature limit ($\gamma_{i\ni}=\Gamma_{i\ni}=\gamma/2$) one has $\braket{P_{12}}=2/3$ in agreement with Eq.~\eqref{system_corr_3}. So for the total spin we obtain
  \begin{equation}
    \label{eq:Szcorr}
    \braket{S_z^2}=\frac{1}{3}.
  \end{equation}
The correlators define the initial conditions for the time correlation functions.

Then the set of the correlators of $\delta S_z(t)$ with the other operators taken at time $t+\tau$ obey the same equations of motion, Eqs.~\eqref{system_spin} and~\eqref{system_corr_3}, for $\tau>0$~\cite{ll3_eng}. Moreover, the spin autocorrelation function is an even function of $\tau$, which allows us to find $\braket{\delta S_z(t)\delta S_z(t+\tau)}$ and the spin noise spectrum $(\delta S_z^2)_\omega$ after Eq.~\eqref{eq:spectrum}. We note, that the spin noise spectrum can be also calculated directly in the frequency domain replacing the time derivatives in equations of motion with the multipliers $-\i\omega$~\cite{noise-excitons}.

The spin noise spectrum is shown in Fig.~\ref{figure2}(b) for the same system parameters as in the panel (a). It again consists of two contributions. (i) A peak at frequency $\omega\sim\delta$, which corresponds to the spin precession in the Overhauser field~\cite{NoiseGlazov,noise-CPT}. Its shape reproduces the distribution function of the absolute values of the Overhauser field~\cite{NoiseGlazov,PolarizedNuclei}. (ii) A peak at zero frequency, which corresponds to the slow spin decay at long times. This peak shows the divergence $(\delta S_z^2)_\omega\propto\ln(1/\omega)$ at $\omega\to0$ in agreement with the asymptotic $\braket{\delta S_z(t)\delta S_z(t+\tau)}\propto 1/\tau$ in the time domain. The logarithmic asymptote for the spin noise spectrum is shown in the inset in Fig.~\ref{figure2}(b).

Thus the spin relaxation and the spin noise spectrum essentially describe the same spin dynamics in the time and frequency domains, respectively.


\section{Limiting cases}
\label{sec:limits}

The main result of the previous section is the very slow power law decay $\propto 1/t$ of the spin polarization despite all the necessary ingredients for the spin relaxation in the model. This result corresponds to the divergence of the spin noise spectrum at zero frequency $\propto\ln(1/\omega)$. In this section we derive these asymptotes in limiting cases, when one of the system parameters $\delta$, $J/\hbar$, or $\gamma$ is much larger than the two others.

\subsection{Strong hyperfine interaction}

The limit $\delta\gg J/\hbar,\gamma$ corresponds to the two nearly independent QDs, where the spins $\bm s_{1,2}$ precess around the corresponding nuclear fields $\bm \Omega_{1,2}$. As a result of this precession the initial spin polarization on average decays three times on the timescale $\sim1/\delta$~\cite{schulten}. The one third of spin polarization on average is parallel to the static fluctuation of the Overhauser field and does not decay at this timescale. The exchange interaction only slightly changes the eigenfunctions and does not lead to the complete spin relaxation. By contrast, the hopping, being incoherent process, leads to the complete decay of the spin polarization. As a result the exchange interaction in this limit can be neglected, while the hopping can not.

\begin{figure*}[t]
  \centering
  \includegraphics[width=\linewidth]{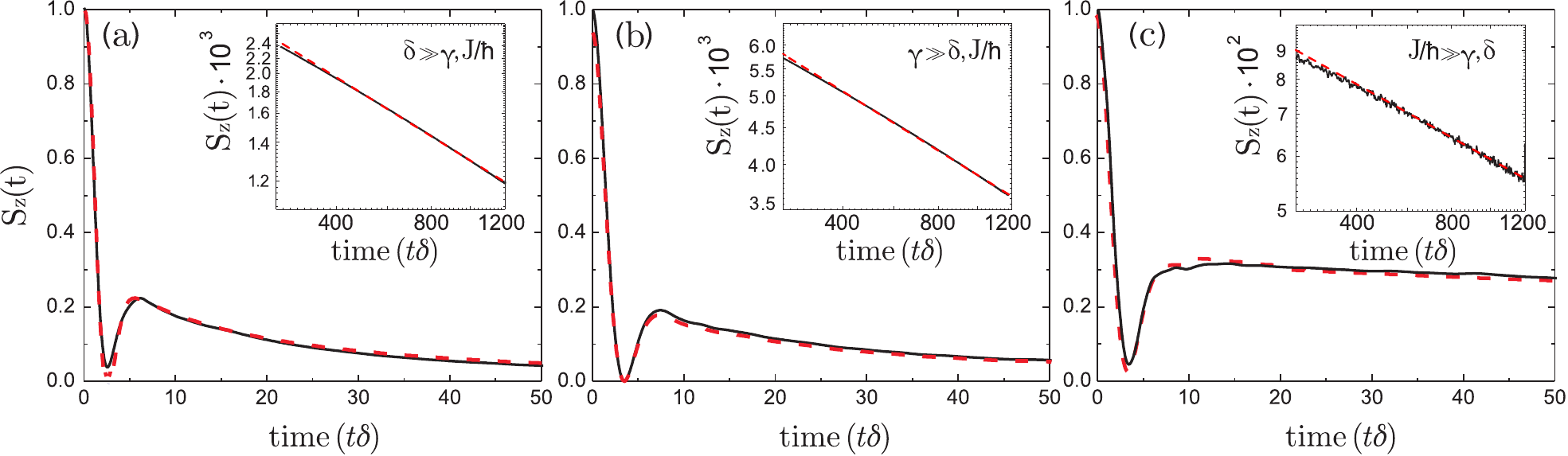}
  \caption{Relaxation of the spin polarization $S_z(t)$ calculated numerically (black solid curves) and analytically (red dashed curves) for the three limiting cases (a) $J=0.04\hbar\delta$ and $\gamma=0.2\delta$ [Eq.~\eqref{eq:lim_delta}], (b) $J=0.2\hbar\delta$ and $\gamma=5\delta$ [Eq.~\eqref{eq:lim_gamma}], and (c) $J=5\hbar\delta$ and $\gamma=0.2\delta$ [Eq.~\eqref{eq:lim_J}]. The initial conditions are $\bm s_{1,2}=\bm e_z/2$. The insets show the power law decay $\propto 1/t$ for the same parameters.}
  \label{fig:limits}
\end{figure*}

The spin dynamics in this limit can be described by
Eqs.~\eqref{system_spin} with $J=0$:
  \begin{equation}
    \label{eq:dyn_delta}
    \dot{\bm s}_{i}=\bm \Omega_{i}\times \bm s_{i}-\frac{\gamma}{2}\cdot(\bm s_{i}-\bm s_{\ni}).
  \end{equation}
One  can separate the spin components parallel and perpendicular to the nuclear field as
\begin{equation}
  s_{i\parallel}=\bm n_i\bm s_i,
  \qquad
  \bm s_{i\perp}=\bm s_i-\bm n_i s_{i\parallel},
\end{equation}
where $\bm n_i=\bm\Omega_i/\Omega_i$. These components approximately obey~\cite{Shumilin2015}
\begin{subequations}
  \begin{equation}
    \dot{\bm s}_{i\perp}=\bm\Omega_i\times\bm s_{i,\perp}-\frac{\gamma}{2}\bm s_{i,\perp},
  \end{equation}
  \begin{equation}
    \dot{s}_{i\parallel}=-\frac{\gamma}{2}\left(s_{i\parallel}-s_{\ni\parallel}\cos\theta\right),
  \end{equation}
\end{subequations}
where $\cos\theta=\bm n_1\bm n_2$. The solution of these equations gives
\begin{multline}
  \bm s_{1}(t)+\bm s_{2}(t)=\sum_i\left[\bm s_{i\perp}(0)\cos(\Omega_i t)+\bm n_i\times\bm s_i(0)\sin(\Omega_i t)\right]\\
  +\frac{\bm n_1+\bm n_2}{2}\left[s_{1\parallel}(0)+s_{2\parallel}(0)\right]\e^{-t\gamma(1-\cos\theta)/2}\\
  +\frac{\bm n_1-\bm n_2}{2}\left[s_{1\parallel}(0)-s_{2\parallel}(0)\right]\e^{-t\gamma(1+\cos\theta)/2}.
  \label{eq:s12}
\end{multline}
This expression should be averaged over the distribution of $\bm \Omega_i$ [see Eq.~\eqref{eq:F}]:
\begin{multline}
  \label{eq:lim_delta}
  \braket{\bm s_{1}(t)+\bm s_{2}(t)}=\left[\bm s_1(0)+\bm s_2(0)\right]\\
  \times\frac{2}{3}\left\lbrace\left[1-\frac{\left(\delta t\right)^2}{2}\right]\e^{-\left(\delta t\right)^2/4}+\frac{\e^{-\gamma t}+\gamma t-1}{\left(\gamma t\right)^2}\right\rbrace.
\end{multline}
This expression is shown by the red dashed curve in
Fig.~\ref{fig:limits}(a) and agrees with the numerical calculations,
shown by the black solid curve. At $t\gg1/\gamma$ this expression
yields the power law decay~\footnote{This qualitative result can not be obtained in the approximation of hopping with some average spin independent rate~\cite{Taylor2007}.}
\begin{equation}
  \braket{\bm s_{1}(t)+\bm s_{2}(t)}=\frac{2}{3\gamma t}.
\end{equation}
This expression is shown by the red dashed line in the inset in Fig.~\ref{fig:limits}(a).

Since the spin correlation functions also obey the equation like Eqs.~\eqref{eq:dyn_delta}, the spin noise spectrum can be found simply as a Fourier transform of Eq.~\eqref{eq:lim_delta} with $\bm s_1(0)+\bm s_2(0)=\bm e_z/3$ [see Eq.~\eqref{eq:Szcorr}]:
\begin{equation}
  \label{eq:spec_delta}
  \left(S_z^2\right)_\omega=\frac{1}{\delta}f\left(\frac{\omega}{\delta}\right)+\frac{1}{\gamma}g\left(\frac{\omega}{\gamma}\right),
\end{equation}
where we introduce the functions
\begin{equation}
  f(x)=\frac{8}{9}\sqrt{\pi}x^2\e^{-x^2},
\end{equation}
\begin{equation}
  g(x)=\frac{2}{9}\left[\pi|x|+\ln\left(1+1/x^2\right)-2\left(1+x\arctg x\right)\right].
\end{equation}
The analytical expression for the spin noise spectrum in this limit is shown in Fig.~\ref{fig:spectra} by the blue dashed curve and agrees with the numerical calculations (blue solid curve).
At low frequencies the spin noise spectrum diverges as
\begin{equation}
  \left(S_z^2\right)_\omega=\frac{4}{9\gamma}\ln\left(\frac{\gamma}{\omega}\right),
\end{equation}
as expected.

\subsection{Fast hopping between QDs}

In the limit $\gamma\gg\delta,J/\hbar$ one could expect, that the spin polarization quickly decays to zero because of the fast hops of electrons into one QD, where the total spin is zero. This, however does not happen, because of the spin blockade: when the two spins are parallel to each other, the electrons do not hop.

It is convenient to rewrite Eqs.~\eqref{eq:spin_i} as
\begin{subequations}
  \begin{equation}
    \label{eq:S}
    \dot{\bm S}=\bm\Omega\times\bm S+\Delta\bm\Omega\times\Delta\bm S,
  \end{equation}
  \begin{equation}
    \Delta\dot{\bm S}=\Delta\bm\Omega\times\bm S-\gamma\Delta\bm S-2\frac{J}{\hbar}\bm s_1\times\bm s_2,
  \end{equation}
\end{subequations}
where $\Delta\bm S=\bm s_1-\bm s_2$, $\bm\Omega=(\bm\Omega_1+\bm\Omega_2)/2$ and $\Delta\bm\Omega=(\bm\Omega_1-\bm\Omega_2)/2$. In the lowest order in $J/(\hbar\gamma)$ the term with $\bm s_1\times\bm s_2$ can be neglected, while $\Delta\bm S$ in the second equation quickly relaxes to the value
\begin{equation}
  \Delta\bm S=\frac{\Delta\bm\Omega\times\bm S}{\gamma}.
\end{equation}
From Eq.~\eqref{eq:S} one can see, that $\bm S$ precesses around $\bm\Omega$, while its projection on $\bm\Omega$ decays due to the second term. Therefore one can solve separately equation for these two components and find
\begin{multline}
  \braket{\bm S(t)}=\bm S(0)\\
  \times\left\langle\sin^2(\theta)\cos\left(\Omega t\right)+\cos^2(\theta)\exp\left(-\frac{\Delta\Omega^2\sin^2(\theta')}{\gamma}t\right)\right\rangle,
\end{multline}
where the angular brackets denote averaging over $\bm\Omega_{1,2}$, $\theta$ is the angle between $\bm\Omega$ and $\bm S(0)$, and $\theta'$ is the angle between $\Delta\bm\Omega$ and $\bm\Omega$.

\begin{figure}[t]
  \centering
  \includegraphics[width=\linewidth]{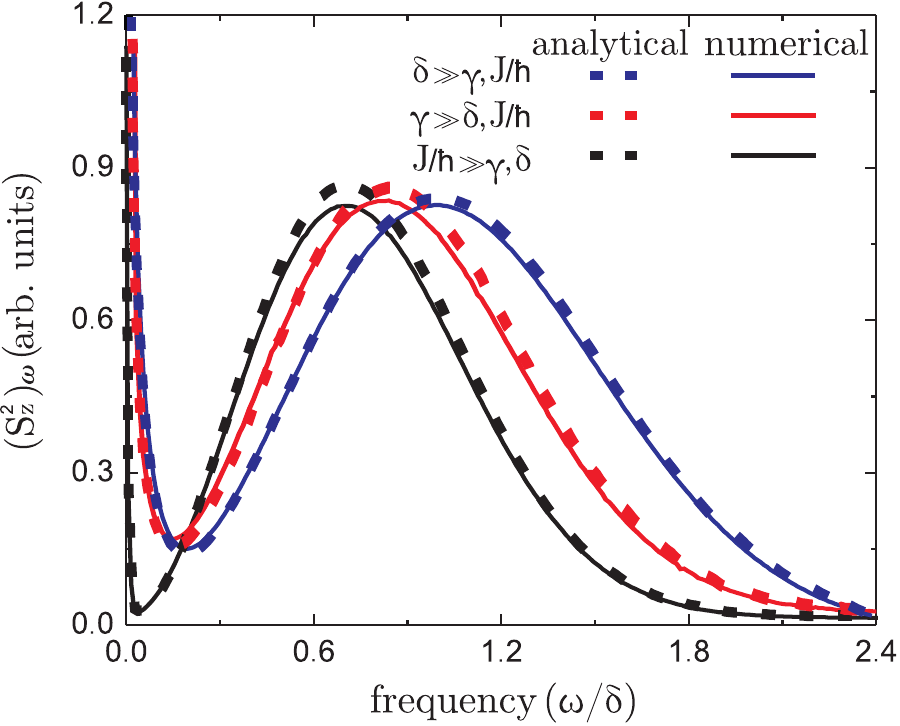}
  \caption{Spin noise spectra calculated numerically (solid curves) and analytically (dashed curves) for the same parameters as in Fig.~\ref{fig:limits}(a) (blue curves), (b) (red curves), and (c) (black curves), see Eqs.~\eqref{eq:spec_delta}, \eqref{eq:spec_gamma}, and~\eqref{eq:spec_J}, respectively.}
  \label{fig:spectra}
\end{figure}

The frequencies $\bm\Omega$ and $\Delta\bm\Omega$ are normally distributed, similarly to Eq.~\eqref{eq:F}, but with $\delta$ being $\sqrt{2}$ times smaller. This allows us to find ultimately
\begin{equation}
  \label{eq:lim_gamma}
  \braket{\bm S(t)}=\bm S(0)\frac{2}{3}\left[\left(1-\frac{(\delta t)^2}{4}\right)\e^{-(\delta t)^2/8}+\frac{\gamma}{2\gamma+\delta^2 t}\right].
\end{equation}
This expression is plotted in Fig.~\ref{fig:limits}(b). One can see, that it is very similar to the panel (a) despite the opposite relation between the parameters. At long time scales the spin polarization decays as
\begin{equation}
  \braket{\bm S(t)}=\bm S(0)\frac{2\gamma}{3\delta^2 t}
\end{equation}
in agreement with the general result of the previous section.

Similarly to the previous subsection the spin noise spectrum can be derived simply performing the Fourier transform of Eq.~\eqref{eq:lim_gamma}:
\begin{equation}
  \label{eq:spec_gamma}
  \left(S_z^2\right)_\omega=\frac{\sqrt{2}}{\delta}f\left(\frac{\sqrt{2}\omega}{\delta}\right)+\frac{\gamma}{\delta^2}h\left(\frac{2\gamma\omega}{\delta^2}\right),
\end{equation}
where we introduced
\begin{equation}
  h(x)=\frac{2}{9}\left\lbrace\sin(|x|)\left[\pi-2\Si(|x|)\right]-2\cos(x)\Ci(x)\right\rbrace,
\end{equation}
with $\Si(x)$ and $\Ci(x)$ being the sine and cosine integral functions, respectively. This expression is shown by the red dashed curve in Fig.~\ref{fig:spectra} and agrees with numerical calculations. At low frequencies one finds
\begin{equation}
  \left(S_z^2\right)_\omega=\frac{4\gamma}{9\delta^2}\ln\left(\frac{\delta^2}{\gamma\omega}\right),
\end{equation}
so the spectrum again diverges logarithmically.

\subsection{Strong exchange interaction}

In the limit $J/\hbar\gg\delta,\gamma$ the spins strongly couple into the triplet and singlet. The hyperfine interaction weakly mixes these states, while the electron hopping is possible only in the singlet state because of the Pauli spin blockade.

The equations for spin dynamics in this limit can be obtained from Eqs.~\eqref{system_spin} in the lowest order in $\hbar\bm\Omega_{1,2}/J$. Note, that the terms containing $\gamma$ again can not be neglected, because they lead to the complete spin decay at long time scales. As a result we obtain
\begin{subequations}
  \begin{equation}
    \label{eq:JS1}
    \dot{\bm S}=\bm\Omega\times\bm S+\Delta\bm\Omega\times\Delta\bm S,
  \end{equation}
  \begin{equation}
    \Delta\dot{\bm S}=\Delta\bm\Omega\times\bm S-\gamma\Delta\bm S-2\frac{J}{\hbar}\bm s_1\times\bm s_2,
  \end{equation}
  \begin{equation}
    (\bm s_1\times\bm s_2)\dot{}=\frac{J}{2\hbar}\Delta\bm S-\gamma\bm s_1\times\bm s_2.
  \end{equation}
\end{subequations}
One can see, that $\Delta\bm S$ and $\bm s_1\times\bm s_2$ decay much faster, than $\bm S$, so the time derivatives in the second and third equations can be set to zero. This gives
\begin{subequations}
  \begin{equation}
    \bm s_1\times\bm s_2=\frac{J}{2\hbar\gamma}\Delta\bm S,
  \end{equation}
  \begin{equation}
    \Delta\bm S=\frac{\hbar^2\gamma}{J^2}\Delta\bm\Omega\times\bm S.
  \end{equation}
\end{subequations}
Substituting the last expression in Eq.~\eqref{eq:JS1} and averaging the solution over the nuclear fields we find
\begin{equation}
  \label{eq:lim_J}
  \braket{\bm S(t)}=\bm S(0)\frac{2}{3}\left[\left(1-\frac{(\delta t)^2}{4}\right)\e^{-(\delta t)^2/8}+\frac{J^2}{2J^2+\hbar^2\gamma\delta^2 t}\right].
\end{equation}
This expression is shown in Fig.~\ref{fig:limits}(c), and one can see, that the spin polarization in this case decays particularly slow. Indeed at long time scales one finds
\begin{equation}
  \braket{\bm S(t)}=\bm S(0)\frac{2J^2}{3\hbar^2\gamma\delta^2t},
\end{equation}
so the prefactor of $1/t$ is parametrically large in the limit under study, $J/\hbar\gg\delta,\gamma$.

The spin noise spectrum in this limit can be calculated similarly to the previous subsections and the result reads
\begin{equation}
  \label{eq:spec_J}
  \left(S_z^2\right)_\omega=\frac{\sqrt{2}}{\delta}f\left(\frac{\sqrt{2}\omega}{\delta}\right)+\frac{J^2}{\hbar^2\gamma\delta^2}h\left(\frac{2J^2\omega}{\hbar^2\gamma\delta^2}\right).
\end{equation}
This expression is shown by the black dashed curve in Fig.~\ref{fig:spectra} and again agrees with the numerical calculations. At low frequencies one finds
\begin{equation}
  \left(S_z^2\right)_\omega=\frac{4J^2}{9\hbar^2\gamma\delta^2}\ln\left(\frac{\hbar^2\gamma\delta^2}{J^2\omega}\right),
\end{equation}
which shows once again, that the spin correlations decay particularly slow in this limit.

\section{Discussion}
\label{sec:discussion}

We demonstrated, that the spin polarization decays as $\propto1/t$ for any relation between the system parameters. Now let us discuss the applicability limits of our model. In typical GaAs based self assembled QDs the characteristic time scales is defined by $1/\delta\sim1$~ns~\cite{A.Greilich09282007,eh_noise}. At longer time scales a few mechanisms of the spin relaxation can come into play, which can limit the applicability of our model.

A zero magnetic field the on-site electron spin flip-flops due to the electron-phonon and spin-orbit interactions have very low rates because of, e.g., the zero phonon density of states with zero energy. Indeed, according to Refs.~\onlinecite{Elzerman2004,Kroutvar2004,Heiss2005,Amasha2008,Hayes2009,Lu2010} the spin relaxation caused by the direct spin-phonon coupling~\cite{PhysRevB.64.125316,PhysRevB.94.125401} or spin admixture mechanisms~\cite{PhysRevB.64.125316,PhysRevB.66.161318} should exceed $1$~s at magnetic field smaller than $1$~T.

The spin-orbit interaction during the hops leads to the spin rotations~\cite{Raikh,KozubPRL}, which is not taken into account by our model. However, this effect can be simply accounted for by the rotation of the coordinate frames for the two QDs in the spin space~\cite{KKavokin-review}.
Still the small random deviations of the electron hopping trajectory from the semiclassical one~\cite{lyubinskiy07} can not be compensated in the same way.

The most probable limitations of our model are the external excitation of the system, e.g. by optical pulses~\cite{greilich2011optical,singleSpin}, and the nuclear spin dynamics. The nuclear spin precession can be caused either by the strain in the QDs and the quadrupole interaction, or by the Knight field created by electrons. These effects take place at the microsecond time scales~\cite{Bulutay2012,Chekhovich2012} and quench $\propto1/t$ asymptotic behaviour. Nevertheless, we assume that our theory will correctly describe the spin dynamics in double QD on the sub microsecond time scales. In particular the spin relaxation should be described by the power law from a few nanoseconds to a few microseconds, e.g. three orders of magnitude in time and frequency domains. 

\section{Conclusion}
\label{sec:conclusion}

To summarize, we studied the spin dynamics in a double QD taking into account the interplay between the hyperfine interaction, exchange interaction and electrons hopping. We demonstrated numerically, that for arbitrary relation between the system parameters the spin relaxation consists of the partial spin dephasing in a random nuclear field and a universal power law decay $\propto1/t$ at large time scales. The spin noise spectrum of the system similarly consists of the two contributions and diverges as $\propto\ln(1/\omega)$ at low frequencies. We proved our results analytically in the limits, when one of the system parameters exceeds by far the others. We believe, that our result will stimulate further experimental investigations of spin relaxation and spin noise in double QD.

\section{Acknowledgments}

We gratefully acknowledge the fruitful discussions with M. M. Glazov.
All numerical calculations were performed under the Russian Science Foundation financial support (RSF No. 18-72-10002). D.S.S. was partially supported by the RF President Grant No. MK-1576.2019.2, Russian Science Foundation grant No. 19-12-00051 and the Basis Foundation.

\appendix
\section{Explicit form of spin dynamics equations}
\label{app:eqs}

We find it useful to rewrite Eqs.~\eqref{system_spin} in an explicit form:

%
\begin{subequations}
 \begin{equation}
   \frac{\d s_{1}^{x}}{\d t} =\Omega_{1}^{y}s_{1}^{z}-\Omega_{1}^{z}s_{1}^{y}+\frac{J}{\hbar}(s_{1}^{z}s_{2}^{y}-s_{1}^{y}s_{2}^{z})-\frac{\gamma}{2}(s_{1}^{x}-s_{2}^{x}),
 \end{equation}
 \begin{equation}
   \frac{\d s_{1}^{y}}{\d t}=\Omega_{1}^{z}s_{1}^{x}-\Omega_{1}^{x}s_{1}^{z}+\frac{J}{\hbar}(s_{1}^{x}s_{2}^{z}-s_{1}^{z}s_{2}^{x})-\frac{\gamma}{2}(s_{1}^{y}-s_{2}^{y}),
 \end{equation}
 \begin{equation}
   \label{eq:s1z}
   \frac{\d s_{1}^{z}}{\d t}=\Omega_{1}^{x}s_{1}^{y}-\Omega_{1}^{y}s_{1}^{x}+\frac{J}{\hbar}(s_{1}^{y}s_{2}^{x}-s_{1}^{x}s_{2}^{y})-\frac{\gamma}{2}(s_{1}^{z}-s_{2}^{z}),
 \end{equation}
 \begin{equation}
  \frac{\d s_{2}^{x}}{\d t} =\Omega_{2}^{y}s_{2}^{z}-\Omega_{2}^{z}s_{2}^{y}+\frac{J}{\hbar}(s_{1}^{y}s_{2}^{z}-s_{1}^{z}s_{2}^{y})-\frac{\gamma}{2}(s_{2}^{x}-s_{1}^{x}),
 \end{equation}
 \begin{equation}
   \frac{\d s_{2}^{y}}{\d t} =\Omega_{2}^{z}s_{2}^{x}-\Omega_{2}^{x}s_{2}^{z}+\frac{J}{\hbar}(s_{1}^{z}s_{2}^{x}-s_{1}^{x}s_{2}^{z})-\frac{\gamma}{2}(s_{2}^{y}-s_{1}^{y}),
 \end{equation}
 \begin{equation}
   \label{eq:s2z}
   \frac{\d s_{2}^{z}}{\d t} =\Omega_{2}^{x}s_{2}^{y}-\Omega_{2}^{y}s_{2}^{x}+\frac{J}{\hbar}(s_{1}^{x}s_{2}^{y}-s_{1}^{y}s_{2}^{x})-\frac{\gamma}{2}(s_{2}^{z}-s_{1}^{z}).
 \end{equation}
\end{subequations}

\begin{subequations}
 \begin{multline}
    \frac{\d}{\d t}(s_{1}^{x}s_{2}^{y})=\Omega_{1}^{y}s_{1}^{z}s_{2}^{y}-\Omega_{1}^{z}s_{1}^{y}s_{2}^{y}+\Omega_{2}^{z}s_{1}^{x}s_{2}^{x}-\Omega_{2}^{x}s_{1}^{x}s_{2}^{z}\\+\frac{J}{4\hbar}(s_{1}^{z}-s_{2}^{z})-\frac{\gamma}{2}(s_{1}^{x}s_{2}^{y}-s_{1}^{y}s_{2}^{x}),
 \end{multline}
 \begin{multline}
  \frac{\d}{\d t}(s_{1}^{x}s_{2}^{z})=\Omega_{1}^{y}s_{1}^{z}s_{2}^{z}-\Omega_{1}^{z}s_{1}^{y}s_{2}^{z}+\Omega_{2}^{x}s_{1}^{x}s_{2}^{y}-\Omega_{2}^{y}s_{1}^{x}s_{2}^{x}\\+\frac{J}{4\hbar}(s_{2}^{y}-s_{1}^{y})-\frac{\gamma}{2}(s_{1}^{x}s_{2}^{z}-s_{1}^{z}s_{2}^{x}),
 \end{multline}
 \begin{multline}
   \frac{\d}{\d t}(s_{1}^{y}s_{2}^{x})=\Omega_{1}^{z}s_{1}^{x}s_{2}^{x}-\Omega_{1}^{x}s_{1}^{z}s_{2}^{x}+\Omega_{2}^{y}s_{1}^{y}s_{2}^{z}-\Omega_{2}^{z}s_{1}^{y}s_{2}^{y}\\+\frac{J}{4\hbar}(s_{2}^{z}-s_{1}^{z})-\frac{\gamma}{2}(s_{1}^{y}s_{2}^{x}-s_{1}^{x}s_{2}^{y}),
 \end{multline}
 \begin{multline}
  \frac{\d}{\d t}(s_{1}^{y}s_{2}^{z})=\Omega_{1}^{z}s_{1}^{x}s_{2}^{z}-\Omega_{1}^{x}s_{1}^{z}s_{2}^{z}+\Omega_{2}^{x}s_{1}^{y}s_{2}^{y}-\Omega_{2}^{y}s_{1}^{y}s_{2}^{x}\\+\frac{J}{4\hbar}(s_{1}^{x}-s_{2}^{x})-\frac{\gamma}{2}(s_{1}^{y}s_{2}^{z}-s_{1}^{z}s_{2}^{y}),
 \end{multline}
 \begin{multline}
  \frac{\d}{\d t}(s_{1}^{z}s_{2}^{x})=\Omega_{1}^{x}s_{1}^{y}s_{2}^{x}-\Omega_{1}^{y}s_{1}^{y}s_{2}^{x}+\Omega_{2}^{y}s_{1}^{z}s_{2}^{z}-\Omega_{2}^{z}s_{1}^{z}s_{2}^{y}\\+\frac{J}{4\hbar}(s_{1}^{y}-s_{2}^{y})-\frac{\gamma}{2}(s_{1}^{z}s_{2}^{x}-s_{1}^{x}s_{2}^{z}),
 \end{multline}
 \begin{multline}
   \frac{\d}{\d t}(s_{1}^{z}s_{2}^{y})=\Omega_{1}^{x}s_{1}^{y}s_{2}^{y}-\Omega_{1}^{y}s_{1}^{x}s_{2}^{y}+\Omega_{2}^{z}s_{1}^{z}s_{2}^{x}-\Omega_{2}^{x}s_{1}^{z}s_{2}^{z}\\+\frac{J}{4\hbar}(s_{2}^{x}-s_{1}^{x})-\frac{\gamma}{2}(s_{1}^{z}s_{2}^{y}-s_{1}^{y}s_{2}^{z}).
 \end{multline}
%
 \begin{equation}
   \frac{\d}{\d t}(s_{1}^{x}s_{2}^{x})=\Omega_{1}^{y}s_{1}^{z}s_{2}^{x}-\Omega_{1}^{z}s_{1}^{y}s_{2}^{x}+\Omega_{2}^{y}s_{1}^{x}s_{2}^{z}-\Omega_{2}^{z}s_{1}^{x}s_{2}^{y}-\frac{\gamma}{4}X,
 \end{equation}
 \begin{equation}
  \frac{\d}{\d t}(s_{1}^{y}s_{2}^{y})=\Omega_{1}^{z}s_{1}^{x}s_{2}^{y}-\Omega_{1}^{x}s_{1}^{z}s_{2}^{y}+\Omega_{2}^{z}s_{1}^{y}s_{2}^{x}-\Omega_{2}^{x}s_{1}^{y}s_{2}^{z}-\frac{\gamma}{4}X,
 \end{equation}
 \begin{equation}
 \frac{\d}{\d t}(s_{1}^{z}s_{2}^{z})=\Omega_{1}^{x}s_{1}^{y}s_{2}^{z}-\Omega_{1}^{y}s_{1}^{x}s_{2}^{z}+\Omega_{2}^{x}s_{1}^{z}s_{2}^{y}-\Omega_{2}^{y}s_{1}^{z}s_{2}^{x}-\frac{\gamma}{4}X,
 \end{equation}
\end{subequations}
%

We recall, that in order to calculate the spin noise spectrum the time derivatives can be replaces with $-\i\omega$, and $1/6$ should be added in the right hand side of Eqs.~\eqref{eq:s1z} and~\eqref{eq:s2z}. Then the double real part of $s_1^z+s_2^z$ equals to the spin noise spectrum $(\delta S_z^2)_\omega$.

\end{document}